\documentstyle[aps,preprint]{revtex}
\begin{document}
\draft
\def\zt{Zlatko Te\v sanovi\' c}
\title{\bf Anomalous Behavior of the Upper Critical Field in Extreme
Type-II Superconductors at Low Temperatures 
 }
\newcommand{\q}{\vec{q}} 
\newcommand{\r}{\vec{r}}
\newcommand{\k}{\vec{k}}
\newcommand{\be}{\begin{equation}}
\newcommand{\ee}{\end{equation}}
\author{
 Sa\v {s}a Dukan and Oskar Vafek\thanks{Present address: Department of Physics and Astronomy, Johns Hopkins University, Baltimore, MD 21218}} 
\address{
\sl  Department of Physics and Astronomy, Goucher College, Baltimore, MD 21204
} 
\maketitle

\begin{abstract}
~~We present a detailed numerical calculation of the upper critical field
$H_{c2}(T)$ for a bulk extreme type-II superconductor.
Particular emphasis is placed on the high-field, low-temperature regime
of the HT-phase diagram. In this regime it is necessary to go beyond the 
standard semi-classical theory and include the effects of Landau quantization 
of the electronic motion on the superconducting state.
The presence of Landau level quantization induces an upward curvature 
in $H_{c2}(T)$ at $\sim 10\%$ of $T_{c0}$ for those superconducting
systems in which the slope of $H_{c2}(T)$ at $T_{c0}$ is $\geq $ 0.2 Tesla/Kelvin. We construct a simple analytical model that can account for this behavior
based on the renormalization of the BCS coupling constant by the off-diagonal
pairing of electrons on Landau levels.
\end{abstract}
\pacs
{PACS numbers: 74.25Dw, 74.60-w, upper critical field, Landau levels, low temperatures\\
Corresponding author: Sa\v{s}a Dukan, Goucher College, 1021 Dulaney Valley
Road, Baltimore, MD 21204\\
Phone: (410)337-6323, Fax: (410)337-6408, email: sdukan@goucher.edu} 
\narrowtext


There has been considerable recent evidence of an anomalous ''divergence"
of the upper critical magnetic field $H_{c2}(T)$ at low temperatures in a 
number of ''low-temperature" high-
temperature superconductors (HTS). Upward curvature and the anomalous 
divergence was observed in Bi$_2$Sr$_2$CuO$_y$ 
\cite{osofsky} and Tl$_2$Ba$_2$CuO$_6$ thin films \cite{mackenzie}; K$_{x}$Ba$_{1-x}$BiO$_3$ single crystals \cite{affronte,russian}; Tl$_2$Mo$_6$Se$_6$ compound \cite{brusetti} as well in borocarbide intermetallic superconductors \cite{naugle}. 
These observations are in contradiction with the standard semi-classical Werthamer-Helfand-Hohenberg (WHH) theory \cite{whh1,whh2} which predicts the saturation of $H_{c2}(T)$ at low temperatures. WHH theory yields a finite value of
the upper critical field at zero temperature, $H_{c2}^{WHH}(0)$, and offers an elegant method for determining this field in clean superconductors as $0.693T_{c0}(-dH_{c2}/dT)_{T=
T_{c0}}$, where $(dH/dT)_{T=T_{c0}}$ is the slope of the upper critical field
evaluated at the zero-field transition temperature $T_{c0}$ (later in the text  called the WHH-slope). For almost three decades, the WHH method has enabled 
experimentalists to determine $H_{c2}(T)$ at low temperatures from the high 
temperature and low magnetic field data (around $T_{c0}$) in conventional 
low-temperature superconductors . Recent advances in obtaining
high magnetic fields in a laboratory environment \cite{bobinger} have revealed 
deviations of the experimental data from WHH theory at low temperatures.

WHH theory is based on the semiclassical phase-integral approximation originally
due to Gor'kov \cite{gorkov} which depends on the following assumption: the bending of the semiclassical paths
of the electrons by the magnetic field is negligible over the range of the 
single-particle Green's function at zero magnetic field. The latter is given
by $v_F/(2\pi k_BT)$, while the radius of a semiclassical path equals $l^2k_F$,
where $v_F$ and $k_F$ are the Fermi velocity and wave vector, and $l=\sqrt{\hbar c/eH}$ is the magnetic length. For a clean superconductor, this assumption 
reads $l^2k_F\gg v_F/(2\pi k_BT)$ or equivalently 
$\hbar \omega _c\ll k_BT$, where $\omega _c=eH/m^*c$ is the cyclotron frequency.
In dirty superconductors (with significant impurity concentration), it translates to  $\hbar \omega _c 
\ll \Gamma $ where $\Gamma$ is the scattering rate due to disorder.
From these conditions, it is clear that the semiclassical WHH theory ignores 
quantum Landau level (LL) effects, an assumption justified at low fields where the electrons occupy a huge number of closely spaced Landau levels (LL's). In this case, the
temperature and/or impurity scattering broaden LL's and reduce the significance 
of LL quantization.
 In the opposite limit of high magnetic fields and low temperatures, where $k_BT
\ll \hbar \omega _c$ and $\Gamma \ll \hbar \omega _c$, the quantization effects are of crucial importance. It was shown by Te\v{s}anovi\'{c} {\it et al.} \cite{zbt} that the discreteness of the
LL's leads to a breakdown
of the semi-classical picture. When the LL level structure is fully accounted for in the BCS theory, one finds that, for a pure case or for a moderate level
of impurities, superconductivity does not terminate above the semiclassical  $H_{c2}^{WHH}(T)$ line, but exhibits reentrant behavior where superconductivity
is enhanced by the magnetic field $H$ (where $H\gg H_{c2}^{WHH}(T)$) \cite{zbt,norman}.
A similar reentrant phase at high fields was recently proposed in organic quasi-one-dimensional superconductors \cite{dupuis}. As a 
consequence of the underlying LL structure $H_{c2}(T)$, or rather $T_c(H)$, develops oscillations near $H_{c2}^{WHH}(0)$ signalling the passage of the LL through the chemical potential. Similar types of quantum oscillations, with the same origin, have been predicted in various other measurable 
quantities which are particularly pronounced in two-dimensional systems \cite{norman,maniv}.
 
The LL quantization of electronic orbits {\em within the superconducting state} is now an experimental fact  in numerous extreme type-II systems.
The strongest evidence so far comes from recent observations 
of de Haas-van Alphen (dHvA) 
oscillations deep in the mixed state of 
various superconducting materials ranging from A-15's 
and boro-carbides to
high-$T_c$ cuprates \cite{book,book1,prl}. What all these materials (some of them also being samples  with a ``diverging" upper critical field) have in common is their
extreme type-II character: the WHH slope in these systems is comparable to $\sim 0.2 $ Tesla/Kelvin
and is often higher \cite{osofsky,mackenzie,affronte,russian,brusetti,naugle}. In such systems the cyclotron 
splitting of LL's near $H_{c2}^{WHH}(0)$, $\hbar \omega _{c2}(0)$, 
where $\omega_{c2}(0)\equiv eH_{c2}(0)/m^*c$, is comparable
to $k_BT_{c0}$   and there is a 
large region in the H-T phase diagram in which the LL structure {\em within} the
superconducting phase is well defined whith $\hbar \omega _c > \Delta (T,H), 
k_BT$ and $\Gamma $ (here $\Delta (T,H)$ is the BCS gap) \cite{note}. The boundaries of 
this high field and low-temperature region in the $H-T$ diagram, 
$H^*$ and $T^*$, extend to fields as low as
$H^*\sim 0.5H_{c2}(0)$ and temperatures as high as $T^*\sim 0.3T_{c0}$ \cite{book1}. In contrast,
the size of this region in conventional type-II superconductors 
(like Nb) is negligible. Within the 
BCS theory, the scale of the cyclotron splitting between LL's near 
$H_{c2}(0)$ in conventional systems is set by the condensation energy, 
$\sim (k_BT_{c0})^2/E_F$, and should be much smaller than either 
the thermal smearing, $\sim k_BT$, or the BCS gap $\Delta (T,H)$. Additional smearing due to the disorder $\Gamma$
makes this high-field and low-temperature region in H-T diagram irrelevant.

In this paper we present a detailed numerical calculation of the upper critical
field $H_{c2}(T)$ for the extreme bulk type-II superconductor with particular emphasis on the high field and low temperature region in the H-T diagram. We
assume a simple isotropic model based on the mean-field theory on LL's developed in
our previous work \cite{book1}. We find LL structure inducing upward curvature in $H_{c2}(T)$ at $\sim 10\%$ of $T_{c0}$ in these systems. In order to account for such behavior, we construct
a scheme based on the renormalization of the BCS coupling constant by the 
off-diagonal pairing of the electrons on LL's. The model presented in this
paper can be used to describe behavior
of a cubic material such as K$_{x}$Ba$_{1-x}$BiO$_3$ \cite{affronte,russian}. 
Furthermore, it can be extended to those anisotropic
systems in which the hopping energy $t^{\perp }$ is larger
than a cyclotron gap at high fileds, {\it i. e.} $t^{\perp }>\hbar \omega _c$.
In such case, an isotropic model, with renormalized masses, correctly captures  high-field behavior of a system. In the opposite limit of a highly anisotropic
material or a quasi-two-dimensional system, mean-field theory is of a limited 
value due to strong fluctuations \cite{zbt}.

Within mean-field theory, the transition line $T_c(H)$, for any field $H$, 
is described as the solution
of the self-consistent equation
\begin{equation}
\Delta ({\bf r})=\frac{V}{\beta}\sum_{\omega } F({\bf r},{\bf r};\omega )\label{self}
\end{equation}
when the amplitude of the superconducting order parameter $\Delta ({\bf r})$ goes to zero. 
$V$ is the usual BCS interaction strength and $\beta =1/k_BT$.  $F({\bf r},{\bf r};\omega )
$ is the anomalous Green's function with  $\omega = 2\pi k_BT(m+1/2)$ being the electron Matzubara frequency (the Matzubara index $m$ is dropped in order not to be confused  with the
LL index later in the text). This anomalous Green's function  is constructed from the wavefunction of the individual electrons in a magnetic field (for review see reference \cite{sasad}). For a clean three-dimensional
(3D) superconductor in high magnetic field equation (\ref{self}) reduces to the form
\begin{equation}
1=\frac{V}{2\pi l^2\beta }\sum_{n,n'=0}^{\infty }\frac{(n+n')!}{n!n'!2^{n+n'}}\int
\frac{dk_z}{8\pi }\sum _{\omega }\frac{1}{\imath \omega -\varepsilon _{n\pm }(k_z)}
\times \frac{1}{-\imath \omega -\varepsilon _{n'\mp }(k_z)}
\label{self1}
\end{equation} 
where $n$ and $n'$ are indices of the Landau levels participating in the superconducting pairing and $k_z$ is the momentum along the field direction.
The electronic energies in the magnetic field
\begin{equation}
\varepsilon _{n\pm }(k_z)=\frac{\hbar ^2\k_z^2}{2m^*}+\hbar \omega _c(n+\frac{1}{2})\mp g\frac{\hbar eH}{4m}-\mu =\frac{\hbar ^2}{2m^*}(k_z^2-k_{Fn\pm}^2)
\label{energy}
\end{equation}
are measured from the chemical potential $\mu $. In the presence of Zeeman
splitting ($g\neq 0$), for each value of $n$ and for each index ``plus" and
``minus" there is a corresponding Fermi momentum  
$k_{Fn\pm }$ determined by the conditions (for $T/\mu \ll 1$)
\begin{eqnarray}
k_{Fn\pm }=\sqrt{\frac{2m^*}{\hbar ^2}(\mu - \hbar \omega _c(n+1/2)\pm g\frac{\hbar eH}{4m})}\\
n_e=\frac{1}{2\pi ^2\l^2}\sum_{n\pm}k_{Fn\pm }
\label{fermi}
\end{eqnarray}
where $n_e$ is the electronic density and the sum is over all real $k_{Fn\pm }$.
In deriving self-consistent equation \ref{self1} we assume that the order
parameter $\Delta ({\bf r})$ has a standard Abrikosov form \cite{abrikosov1}.
In the LL representation Abrikosov order parameter belongs to the
lowest Landau level (LLL) for the Cooper charge $e^*=2e$ having the lowest kinetic energy of the center-of-mass motion of Cooper pairs and therefore
the highest transition temperature. Note 
though, that in two dimensions (2D) special circumstances might arise where the contributions to the order
parameter from the higher LL's become competitive leading to the higher
transition temperature \cite{norman}. However, in 3D, higher LL contributions lead to the 
significantly lower transition temperature \cite{zbt,prb}.  

In the framework of the BCS theory, the density of states
(DOS) of the electrons in a magnetic field exhibits a divergence whenever the bottom of a
LL crosses the chemical potential. As a consequence, strong oscillations in 
$T_{c}(H)$ develop at high fields and low temperature in a clean superconductor \cite{zbt}. These oscillations, in principle, could be
observed in extremely clean samples which is rarely the case in an experimental
setup. Therefore, we consider a more realistic system for which the presence
of the impurities and imperfection (or disorder in general) is taken into  account. We assume that the disorder in the sample leads to isotropic broadening of the LL's, the size of which is measured by $\Gamma =\hbar /2\tau$ , where $1/2\tau$ is the scattering rate
due to the disorder. As long as $\Gamma /\hbar \omega_c\ll 1$ the discreteness of the LL structure is preserved but the quantum oscillations are greatly reduced. In this way we are left only with the task of numerically examining the
overall rising trend of $T_c(H)$ at high fields. The broadening of the
LL can be most easily included into the self-consistent equation (\ref{self1}) by a substitution $i\omega \rightarrow i\omega -\Gamma$ in the Matzubara 
frequencies. 

After integration over the momenta $k_z$ and summation over the Matzubara frequencies (with the Debye frequency
$\Omega $ as a UV cut-off), the
equation (\ref{self1}) can be put in the form
\begin{eqnarray}
\frac{1}{\lambda }=\sum_{n,n'=0}^{\infty }\frac{(n+n')!}{n!n'!2^{n+n'}}\frac
{N_{1CM}(0)}{2\pi l^2N_{3D}(0)}\frac{1}{4}\left[ \Psi (Z_{1\pm}) +\Psi (Z_{1\pm}^*) 
 -\Psi (Z_{2\pm}) -\Psi (Z_{2\pm}^*)\right]
\nonumber\\
Z_{1\pm}=\frac{\Omega}{2\pi T}+i\frac{\hbar
\omega_c(n-n'\pm g/2)}{4\pi T}
\nonumber\\
Z_{2\pm}=\frac{1}{2}+i\frac{\hbar
\omega_c(n-n'\pm g/2)}{4\pi T}
\label{final}
\end{eqnarray}
where for each value of the index $n$ there are two terms,
one for the ``+" and one for the ``-" sign. $\lambda =VN_{3D}(0)$ is the BCS coupling constant with $N_{3D}(0)$ being the
single-spin electronic DOS in zero field calculated at the Fermi energy $E_F$. $\Psi (z)$ is the Digamma function
of the complex variable $z$, and $z^*$ denotes complex conjugation. $N_{1CM}(0)/2\pi l^2$ is the DOS of the
center-of-mass of the Cooper charge at the chemical potential and it 
compares to $N_{3D}(0)$ as
\begin{equation}
\frac{N_{1CM}(0)}{2\pi l^2N_{3D}(0)}=\frac{\Gamma }{2^{5/2}\sqrt{E_F\hbar \omega _c}(C^2+(\frac{\Gamma }{\hbar \omega _c})^2)^{1/2}(\sqrt{C^2+(\frac{\Gamma }{\hbar \omega _c})^2}-C)^{1/2}}
\label{density}
\end{equation}
where 
\begin{equation}
C=\frac{\mu }{\hbar \omega _c}-\frac{n+n'+1}{2}~.
\label{c}
\end{equation}

Our main goal is to numerically solve the self-consistent equation (\ref{final})
for the transition temperature $T_c(H)$ taking realistic values for
the coupling constant $\lambda $ as well as moderate values for the disorder parameter $\Gamma /E_F$ . In particular, we want to incorporate in our
calculation  the crucial observation that the WHH-slope $(-dH_{c2}/dT)_{T=T_{c0}}$ 
reported in the experiments \cite{osofsky,mackenzie,affronte,russian,brusetti,naugle} is always much larger than $\approx 0.2 $ Tesla/K.
Therefore, we  chose a coupling constant $\lambda $ so that the WHH-slope when  expressed in the
dimensionless units $[-d\omega_c/d(k_BT)]_{T=T_{c0}}$ is larger than $0.27$ \cite{note}.

Figure 1 shows the numerical solution of equation (\ref{final}), {\it i.e.} the plot of  
the upper critical field $H_{c2}(T)$ rescaled by $H_{c2}^{WHH}(0)$ {\it vs.} $T_{c}/T_{c0}$ for the model system with coupling constant
$\lambda =0.35$ (full line) and $\lambda =0.4$ (dotted line), the disorder
parameter $\Gamma /E_F=0.025$ and zero Zeeman splitting ($g=0$). 
For the model with $\lambda =0.35$ the WHH-slope in dimensionless units is $[-d\omega_c/d(k_BT)]_{T=T_{c0}}=0.8$ which corresponds to $(-dH_{c2}/dT)_{T=T_{c0}}=0.58$ Tesla/K. When $\lambda =0.4$, the WHH slope in dimensionless units is $1.108$, or $0.813$ Tesla/K in conventional units. The upper
critical field starts to deviate from the one predicted by  
the WHH theory at $T_{c}/T_{c0} \approx 0.1$ exhibiting 
an anomalous divergence at low temperature as predicted by Te\v{s}anovi\'{c}
{\it et al.} \cite{zbt} and recently seen in experiments \cite{osofsky,mackenzie,affronte,russian,brusetti,naugle}. While disorder
completely washes away the quantum oscillations of the upper critical field at
low temperatures for $\lambda =0.35$ case, they are still visible for the
model with the $\lambda =0.4$.
 
In Figure 2 we plot the ratio $\hbar \omega _c/E_F$ (which is proportional to $H_{c2}$, the proportionality constant being set by the material properties
of the particular superconductor) {\it vs.} $T_c/T_{c0}$ for the model system with
$\lambda =0.35$ in order to illustrate the effect of Zeeman splitting ($g\neq 0$) on the
upper critical field. The overall tendency of the Zeeman splitting is to suppress $T_c$ in the region of the 
H-T diagram where $k_BT \leq \hbar \omega _c$, while the transition line is not
affected by Pauli pair breaking in the high temperature-low field portion of
the diagram. An interesting situation arises when effective $g$ factors are
very close to $g=2m/m^*$, $g=4m/m^*$ or any other even integer (we take the effective cyclotron mass $m^*$ of the order of the electron mass $m$ further in the text).  Unlike the odd $g$ factor case ($g=1$ is an example shown in Figure 2), the anomalous divergence of the upper critical field at low
temperature is not destroyed by Zeeman splitting for even integer $g$ factors. The temperature
at which $H_{c2}(T)$ starts to deviate from the WHH line is reduced by only few
percent from the $g=0$ case. This situation can be understood as follows: When
$g=2$ or $g=4$
the Zeeman splitting is equal to the cyclotron splitting making the $n$th
spin-up LL degenerate with the $(n+1)$th (for $g=2$)
or $(n+2)$th (for $g=4$) spin-down level. In these cases the off-diagonal
terms in equation (\ref{final}), describing LL's separated by the cyclotron gaps, become effectively diagonal ({\it i.e.} degenerate) leading to the anomalous divergence in $H_{c2}(T)$ (see the discussion below). When $g$ is an odd
integer or a fraction, the LL spin degeneracy is completely lifted resulting in the suppressed $H_{c2}(T)$ at low temperatures with a  downward curvature.

The upward curvature of the upper critical field $H_{c2}(T)$ at temperatures $T<0.1T_{c0}$ is the consequence of the competing tendencies of diagonal,
 $n=n'$, and off-diagonal, $n\neq n'$, terms in the self-consistent equation (\ref{final}). The diagonal terms correspond to the Cooper pairs
formed by electrons in the same LL while the off-diagonal terms
represent the electronic pairing of the LL's separated by $\hbar \omega _c$ or more.
Only the diagonal terms posses Cooper singularity and therefore lead to
the increasing trend of the transition temperature as a  
function of field \cite{zbt,norman,maniv}. In clean systems, they also produce strong quantum oscillations in $T_c(H)$ for fields $H>H_{c2}^{WHH}(0)$. Ultimately, the diagonal terms in (\ref{final}) lead to the reentrant superconductivity for magnetic fields $H\gg H_{c2}^{WHH}(0)$. This Quantum Limit regime, reached when all electrons in the
system occupy a single, lowest Landau level, was predicted and investigated by Te\v{s}anovi\'{c} {\it et al.} \cite{zbt} but is not of interest
in our calculation since it can be realized only in extremely high fields. The off-diagonal terms in (\ref{final}) start to dominate at lower fields $H\approx 
H_{c2}^{WHH}(0)$ creating a counter-effect to the rapidly decreasing 
diagonal terms. These {\em non-singular} terms lead to the smooth crossover from
the diverging high-field transition line to the low-field WHH line. This crossover results necessarily in the upward curvature of the upper critical
field $H_{c2}(T)$ at low temperatures. Our goal in this paper is to closely
examine this crossover behavior of the off-diagonal terms and to account for 
the resulting upward curvature in a simple analytic model. First, we notice
that the effective role of the off-diagonal terms in the self-consistent equation
(\ref{final}) is to {\em renormalize} the BCS coupling constant $\lambda $ 
into a new, field and temperature dependent, constant $\tilde{\lambda }(H,T)$ 
through the substitution $1/\lambda \rightarrow 1/\lambda -1/\lambda '(H,T)=1/\tilde{\lambda }(H,T)$ where $1/\lambda '(H,T)$ accounts for the 
off-diagonal, $n\neq n'$, terms . The size of the off-diagonal terms in (\ref{final}) grows as the magnetic field is lowered, leading to the effective increase in the renormalized BCS coupling constant $\tilde{\lambda }(H,T)$
rendering a non-zero transition temperature at lower fields. 

With this approach the solution of the self-consistent equation can
be written in analytical form as
\begin{equation}
T_c(H)=1.134\Omega \exp{\left[ -\frac{2\pi l^2}{\tilde{\lambda }(H,T_c)}\left[   \sum_{n=0}^{n_c}\frac{N_{1n}(0)(2n)!}{N_{3D}(0)2^{2n}(n!)^2}\right] ^{-1}\right]}
\label{simple}
\end{equation}
where $N_{1n}(0)$ is the 1D density of states given by (\ref{density}) and 
(\ref{c}), where
$n=n'$. When $2\pi k_BT<\hbar \omega_c $ the contribution of the off-diagonal terms $1/\lambda '(H,T_c)$ to
the renormalized coupling constant can be obtained by systematic expansion
in $2\pi k_BT_c/\hbar \omega _c$. Assuming that the number of the occupied LL's
$n_c=E_F/\hbar \omega _c$ is much larger than one, {\it i.e.} that the system in question is far away from the Quantum Limit regime (in
a typical experimental setup this is indeed the case, since usually $n_c>30$),
and ignoring the quantum oscillations (they are damped by disorder), we 
perform the expansion of the off-diagonal terms up to $(2\pi k_bT_c/\hbar \omega _c)^2$. Then, it is possible to rewrite the equation (\ref{simple}) 
in very simple analytic form as
\begin{equation}
T_c(H)=1.134\Omega \exp{\left[-\frac{4\sqrt{n_c}}{\sqrt{\pi}}\left( \frac{1}{\lambda}-
\frac{1}{\lambda '(H,T_c)}\right) \right]}
\label{simple1}
\end{equation}
where
\begin{eqnarray}
\frac{1}{\lambda '(H,T_c)}=\exp{\left(- \frac{2\sqrt{\pi}}{n_c}\right)}+\exp{\left(-\frac{1}{2\sqrt{\pi n_c}}\right)}\ln{\sqrt{n_c}}+\frac{1}{2n_c}\left(\frac{2\pi k_BT_c}{\hbar \omega _c}\right)
\nonumber\\
+\left(\frac{2}{3\pi \sqrt{n_c}}+\frac{1}{3\pi \sqrt{n_c^3}}-\frac{\pi^2}{6n_c^2}\ln{(n_c)}\right)\left(\frac{2\pi k_BT_c}{\hbar \omega _c}\right)^2 .
\label{off}
\end{eqnarray}
Equation (\ref{simple1}) can be solved by iteration or can be used as
a formula to fit the experimental data. Figure 3 shows  
$T_c(H)$ computed from (\ref{simple}) with $1/\lambda '(H,T_c)$
in (\ref{off}) expanded to leading order
(dotted line) and to second order (dashed line) in $2\pi k_BT_c/\hbar \omega _c$
compared to the exact numerical solution of self-consistent equation
(\ref{final}) for $\lambda =0.35$. The agreement is excellent in the region
where $2\pi k_BT<\hbar \omega _c$, where the expansion (\ref{off}) is valid.
The simple analytic form (\ref{simple1}) accounts very well for the diverging
upper critical field $H_{c2}(T)$ and its anomalous upward curvature.
When $2\pi k_BT=\hbar \omega _c$ this expansion breaks down as indicated
in Figure 3 by the straight line. At low fields and high temperature, where $2\pi k_BT>\hbar \omega _c$, there is a large deviation of $T_c(H)$ from the exact numerical
solution of (\ref{final}), signalling the breakdown of the expansion. 

To conclude we have presented a detailed numerical calculation of the upper
critical field $H_{c2}(T)$ for a three dimensional extreme type-II superconductor characterized by a large WHH slope. We find that
the Landau level quantization induces an upward curvature in $H_{c2}(T)$
at temperatures $\sim 0.1T_{c0}$. We account for this behavior through
renormalization of the BCS coupling constant $\lambda $ by the off-diagonal pairing of the electrons in the Landau levels. Our work, based on the simple
BCS model, reproduces qualitatively observations of the anomalous
behavior of the upper critical field in the experiments \cite{affronte,russian}, but it cannot account quantitatively for the large deviations in $H_{c2}(T)$ from the WHH line for
temperatures higher than predicted by our theory. We believe 
that extension of this work to a more realistic strong-coupling model 
within the Landau level framework will improve agreement with the
experimental results.

We are grateful to Prof. Zlatko Te\v{s}anovi\'{c} for many useful discussions.
This work has been supported by NSF Research Opportunity Award DMR-9415549, Amendment No. 003.

\newpage
\begin{figure}
\caption{Plot of the upper critical field $H_{c2}$ computed from (6) vs. $T_c/T_{c0}$ for a 3D superconductor  with the BCS coupling constant $\lambda =0.35$
and $\lambda =0.40$ and with no Zeeman splitting. $H_{c2}$ is rescaled by $H_{c2}^{WHH}(0)=0.693T_{c0}(-dH_{c2}/dT)_{T=
T_{c0}}$.}
\label{plot0}
\end{figure}
\begin{figure}
\caption{Upper critical field $H_{c2}\sim \hbar \omega _c/E_F$ for
a 3D superconductor with non-zero Zeeman splitting.} 
\label{plott1}
\end{figure}
\begin{figure}
\caption{Comparison of the exact numerical solution of (6) (full line)  and the solution
of (9) obtained by the expansion of the off-diagonal terms in (6) up to
the leading (dotted line) and the second order (dashed line) in $2\pi k_BT_c/
\hbar \omega _c)$. To the left of the  straight line $2\pi k_BT_c=\hbar \omega _c$ this expansion is valid.} 
\label{plott2}
\end{figure}

\begin{thebibliography}{99}
\bibitem{osofsky} M. S. Osofsky et al., Phys. Rev. Lett. {\bf 71}, 2315 (1993).
\bibitem{mackenzie} A. P. Mackenzie et al., Phys. Rev. Lett. {\bf 71}, 1238 (1993).
\bibitem{affronte} M. Affroante et al., Phys. Rev. B {\bf 49}, 3502 (1994).
\bibitem{russian} V. F. Gantmakher et al., Phys. Rev. B {\bf 54}, 6133 (1996).
\bibitem{brusetti} R. Brusetti et al., Phys. Rev. B {\bf 49}, 8931 (1994).
\bibitem{naugle} K. D. D. Rathnayaka et al., Phys. Rev. B {\bf 55}, (1997)
\bibitem{whh1} E. Helfand and N. R. Werthamer, Phys. Rev. Lett. {\bf 13}, 686 (1964).
\bibitem{whh2} N. R. Werthamer et al., Phys. Rev. {\bf 147}, 295 (1966).
\bibitem{bobinger} G. Boebinger, Physics Today {\bf 42}, 36 (1996).
\bibitem{gorkov} L. P. Gorkov, J. Exp. Theroret. Phys. {\bf 34}, 735 (1958), 
Soviet. Phys. JETP {\bf 7}, 505 (1958).
\bibitem{zbt}Zlatko Te\v {s}anovi\' {c}, M. Rasolt and Li Xing, Phys. Rev.
Lett. {\bf 63}, 2425 (1989); Zlatko Te\v {s}anovi\' {c}, M. Rasolt and L. Xing, Phys. Rev. B {\bf 43}, 288 (1991); M. Rasolt and Z. Te\v {s}anovi\' {c}, Rev.
Mod. Phys. {\bf 64}, 709 (1992).
\bibitem{norman} H. Akera, A. H. MacDonald, S. M. Girvin and M. R. Norman,
Phys. Rev. Lett. {\bf 67}, 2375 (1991); A. H. MacDonald, H. Akera and M. R. 
Norman, Phys. Rev. B {\bf 45}, 10147 (1992); M. R. Norman, H. Akera and 
A. H. MacDonald, Physica C {\bf 196}, 43 (1992).
\bibitem{dupuis}N. Dupuis, G. Montambaux and C. A. R. S\'{a} de Melo, Phys. Rev. Lett. {\bf 70}, 2163 (1993); N. Dupuis and G. Montambaux, Phys. Rev. B {\bf 49}, 8993 (1994).
\bibitem{maniv} T. Maniv, A. I. Rom, I. D. Vagner and P. Wyder, Phys. Rev. 
B {\bf 46}, 8360 (1992).
\bibitem{book} For a review on the dHvA experiments see T. J. B. M. Janssen and M. Springford in "The Superconducting State in Magnetic Fields: Special Topics
and New Trends", Edited by Carlos A. R. Sa de Melo, Series on Directions in 
Condensed Matter Physics, {\bf 13}, 175, World Scientific, Singapore, 1998.
\bibitem{book1} S. Dukan and Z. Te\v{s}anovi\'{c} in "The Superconducting State in Magnetic Fields: Special Topics
and New Trends", Edited by Carlos A. R. Sa de Melo, Series on Directions in 
Condensed Matter Physics, {\bf 13}, 197, World Scientific, Singapore, 1998.
\bibitem{prl} S. Dukan and Z. Te\v{s}anovi\'{c}, Phys. Rev. Lett. {\bf 74},
2311 (1995).
\bibitem{note} Note that for the effective mass, 
$m^*$, of the order of the electron mass, $\hbar \omega _c\sim 1.34$ 
Kelvin at $H=$ 1 Tesla. 
\bibitem{sasad}  S. Dukan and Z. Te\v{s}anovi\'{c}, Phys. Rev. B {\bf 56}, 838 (1997).
\bibitem{abrikosov1} A. A. Abrikosov, JETP {\bf 5} 1174 (1957); Zh. Exp. Ter.
Fiz. {\bf 32}, 1442 (1957).
\bibitem{prb} S. Dukan and Z. Te\v{s}anovi\'{c}, Phys. Rev. B {\bf 49}, 13017 (1994).
\end{thebibliography}
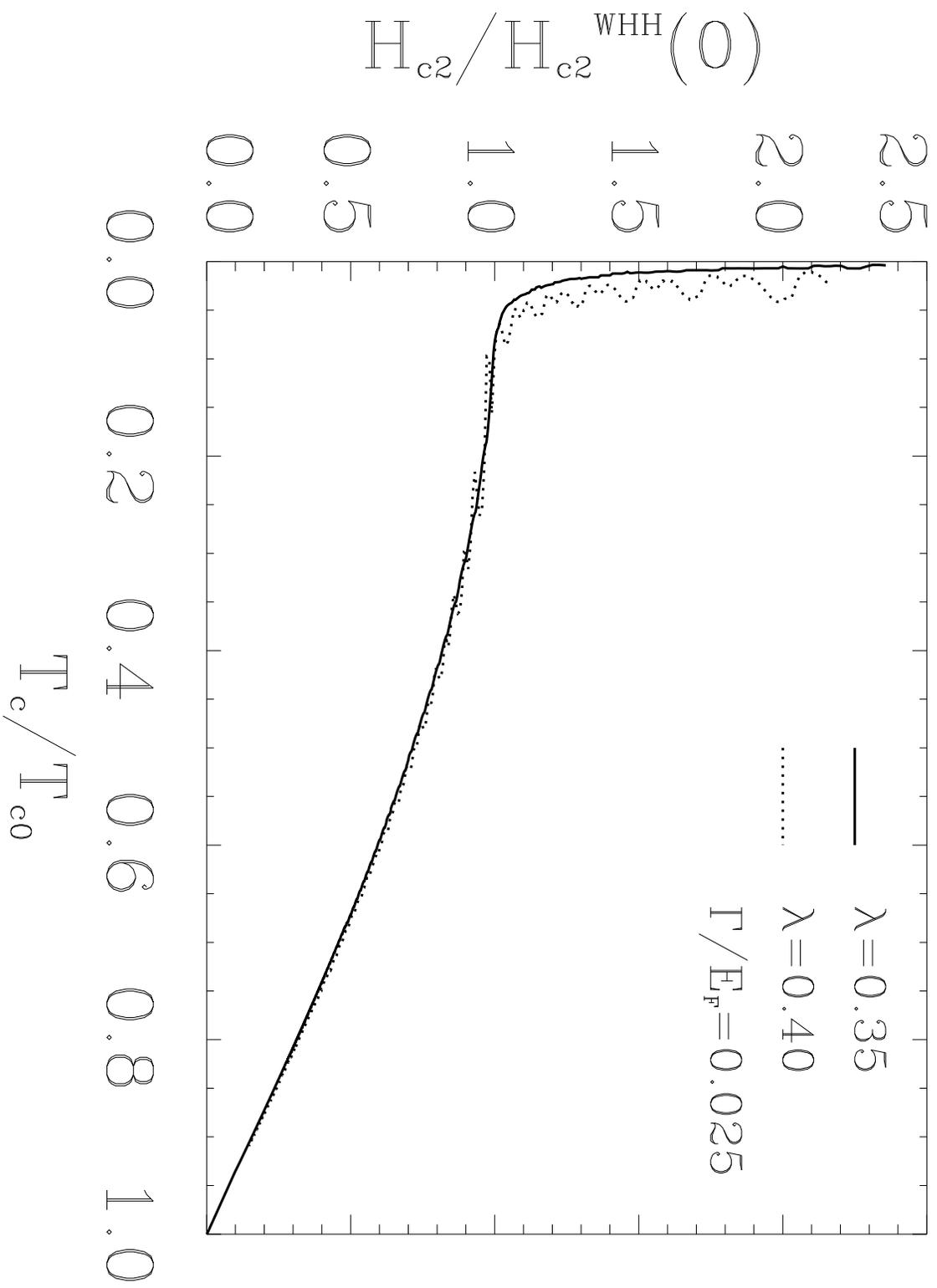
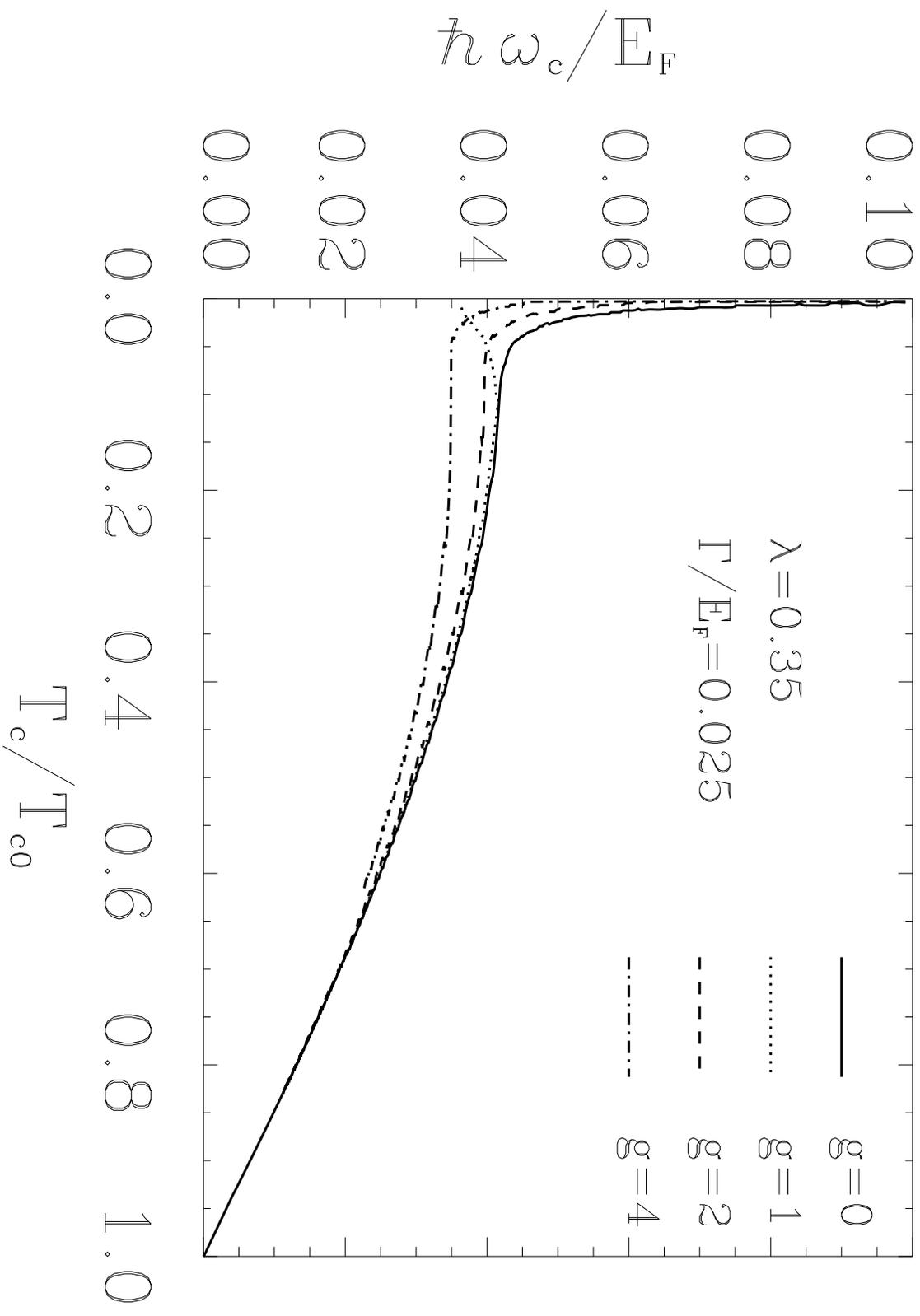
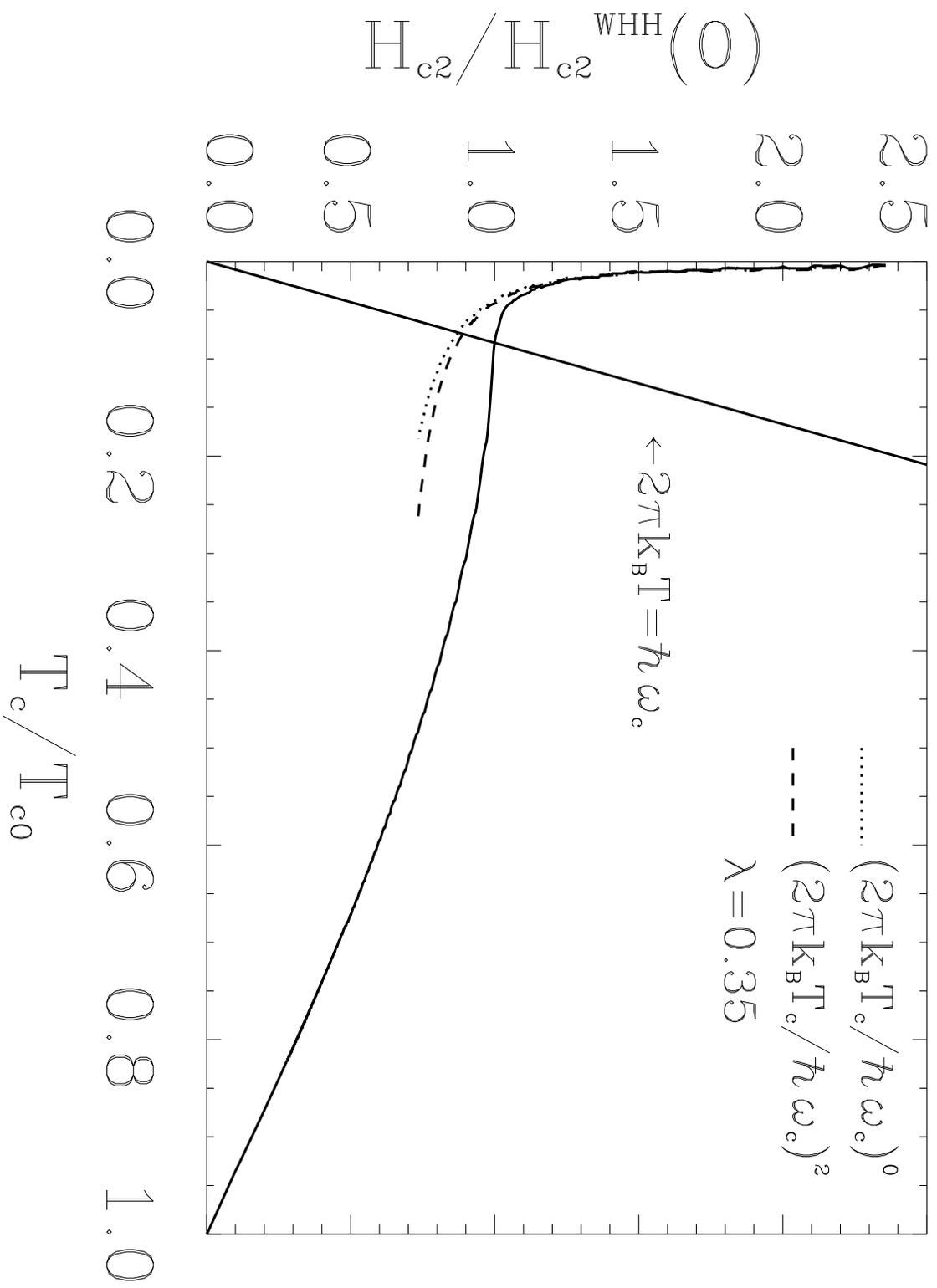
\end{document}